# Magnon-phonon interactions in magnon spintronics

D.A. Bozhko[1], V.I. Vasyuchka[2], A.V. Chumak[3], and A.A. Serga[2]

[1]*Department of Physics and Energy Science, University of Colorado at Colorado Springs, Colorado Springs CO 80918, USA*

*E-mail: dbozhko@uccs.edu*

[2]*Fachbereich Physik and Landesforschungszentrum OPTIMAS, Technische Universität Kaiserslautern, D-67663 Kaiserslautern, Germany*

[3]*Faculty of Physics, University of Vienna, A-1090 Vienna, Austria*

Nowadays, the interaction between phonon and magnon subsystems of a magnetic medium is a hot topic of research. The complexity of phonon and magnon spectra, the existence of both bulk and surface modes, the quantization effects, and the dependence of magnon properties on applied magnetic field, make this field very complex and intriguing. Moreover, the recent advances in the fields of spin-caloritronics and magnon spintronics as well as the observation of the spin Seebeck effect (SSE) in magnetic insulators points on the crucial role of magnons in spin-caloric transport processes. In this review, we collect the variety of different studies in which magnon-phonon interaction play important role. The scope of the paper covers the wide range of phenomena starting from the interaction of the coherent magnons with surface acoustic wave and finishing with the formation of magnon supercurrents in the thermal gradients.

1.   **Introduction**
2.   **Interaction of coherent magnons with phonons**
     2.1.   Bragg scattering of coherent spin waves from surface acoustic waves
     2.2.   Magnon transformations in thermal gradients
     2.3.   Control of anisotropic spin waves propagation by magnetization gradients
     2.4.   Heat-induced magnonic crystal

## 1.   Introduction

A spin wave is a collective excitation of the electron spin system in a magnetic solid [1] and magnon is a spin-wave quantum. Spin-wave characteristics can be varied by a wide range of parameters including the choice of the magnetic material, the shape of the sample as well as the orientation and size of the applied biasing magnetic field [2, 3]. This, in combination with a rich choice of linear and non-linear spin-wave properties [4], renders spin waves excellent objects for the studies of general wave physics. One- and two-dimensional soliton formation [5, 6], non-diffractive spin-wave caustic beams [7-10], wave-front reversals [11, 12], and room temperature Bose–Einstein condensation of magnons [13-16] is just a small selection of examples. On the other hand, spin waves in the GHz frequency range are of large interest for applications in telecommunication systems and radars. Since the spin-wave wavelengths are orders of magnitude smaller compared to electromagnetic waves of the same frequency, they allow for the design of micro- and nano-sized devices for analog data processing (e.g. filters, delay lines, phase shifters, isolators – see e.g. Refs. [17-21]). Nowadays, spin waves and their quanta, magnons, are attracting much attention as data carriers in novel computing devices [22-26]. The field of science that refers to information transport and processing by spin waves is known as *magnonics* [4, 22, 27]. The utilization of magnonic approaches in the field of spintronics, hitherto addressing electron-based spin currents, gave birth to the field of *magnon spintronics* [23, 24]. Magnon spintronics comprises magnon-based elements operating with analog and digital data as well as converters between the magnon subsystem and electron-based spin and charge currents. Moreover, phonons mediate interconversions between magnon and electron sub-systems and, thus, play a crucial role in magnon spintronics.

The current review addresses a selection of topics that form the basis of the understanding of the role of phonons and magnon-phonon interactions in the field of magnon spintronics. First, the interaction of coherent spin waves with coherent acoustic modes [28] is discussed briefly. Second, the manipulation and guiding of the coherent spin waves by the use of heat is discussed [29-32] as well as the conversation of magnon energy into heat [33]. The next chapter addresses the emergent field of spin caloritronics [34] which investigates the interplay between spin- and heat-based transport phenomena. The final chapter deals with the fascinating phenomena of Bose-Einstein condensation of magnons [13-15] and concentrates on the role of phonon modes and heat in them. The review is finished by the brief concluding remarks.

## 2. Interaction of coherent magnons with phonons

### 2.1. Bragg scattering of coherent spin waves from surface acoustic waves

Magnonic crystals (MCs) are artificial magnetic media with periodic variation of their magnetic properties in space. Similarly to photonic crystals operating with light, the Bragg scattering affects the SW spectrum in such a periodic structure and leads to the formation of band gaps – frequencies, at which the SW propagation is prohibited. MCs use the wave nature of magnons to obtain magnon propagation characteristics that are inaccessible by any other means (see reviews [4, 35-38]). This allows one to combine one- and two-dimensional magnon waveguidings with the magnon-based data processing [38]. Furthermore, the reconfigurable [37] and dynamic [38, 39] MCs demonstrate a potential for the realization of universal data processing units, which can perform different operations with data on demand.

Moving magnonic crystals represent a special class of dynamic and tunable crystals utilizing moving Bragg gratings. Such magnonic crystals have been created via a periodic strain induced by a surface acoustic wave (SAW) traveling along a YIG spin-wave waveguide as shown in Figure 1 [28]. Spin-wave scattering in such a crystal differs conceptually from the previously discussed cases since it takes place with a shift in the spin-wave frequency due to the Doppler effect. In fact, a combination of Bragg scattering with the Doppler shift took place in the experiments [28]. The schematic of such scattering is shown in the right panel of Figure 1. One can see that in order to fulfill the conservation laws, the magnon scattering should occur with a shift in the frequency equal to the frequency of the SAW. The wavelengths of the spin waves and SAW should be comparable in order to satisfy momentum conservation.

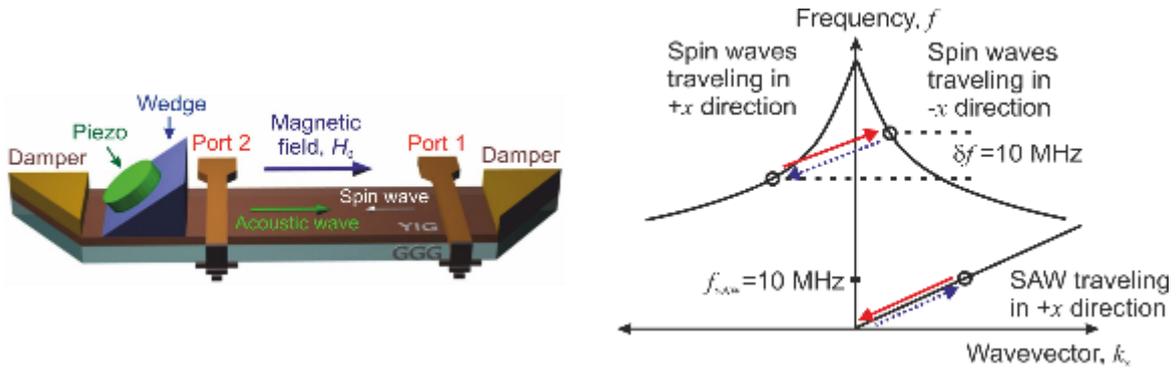

**Figure 1. Scattering of spin waves on surface acoustic wave.** Left panel. Schematic of the moving SAW-induced magnonic crystal. The spin waves are excited and detected in the YIG film by stripline antennas. The surface acoustic wave is excited on the YIG/GGG sample by a piezoelectric quartz crystal and an acrylic wedge transducer [28]. Right panel: Schematic of the dispersion curves for the BVMSW and SAW. Circles indicate waves that participate in Bragg scattering.

The experiment was performed using BVMSWs characterized by negative group velocity [2, 3]. As a result, spin waves scattered from an approaching grating were found to be shifted down in frequency, demonstrating the reverse Doppler effect [40]. In contrast to [40], here the reflection occurred from a crystal lattice rather than from a single reflecting surface and, thus, the wavenumber of the scattered wave was strictly determined by the law of conservation of momentum. Due to this fact, the frequency-shifted wave appeared as a single narrow peak in the transmission characteristic of the magnonic crystal. Apart from such interesting physics, these systems show potential for sensors and signal-processing applications.

### 2.2. Magnon transformations in thermal gradients

In order to investigate the behavior of spin waves propagating in a region of inhomogeneous temperature, a thermal gradient has been applied to a Yttrium Iron Garnet (YIG) waveguide so that spin waves are excited in a hot region near the antenna and propagate into the colder region of the YIG stripe. This situation is shown in the thermal image of Fig. 2a. Using phase-resolved Brillouin light scattering (BLS) spectroscopy the intensity pattern of the interference between the spin-wave scattered light and separately prepared reference light with a constant reference phase has been recorded. Mapping the spatial distribution of the interference signal allows for a direct visualization of the spin-wave wavelength.

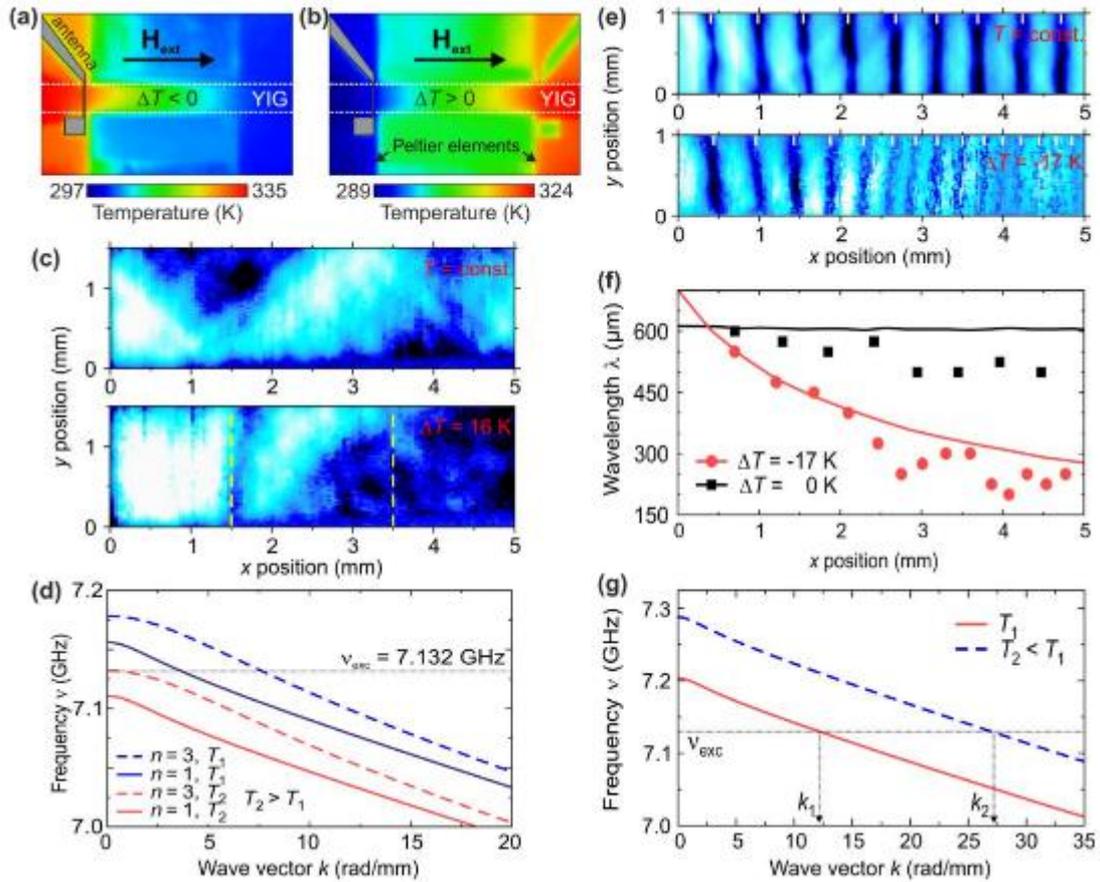

**Figure 2. Spin-wave propagation and transformation in a thermal gradient [29].** Thermal images of the sample with overlaid schematic setup. A microwave antenna excites spin waves in the YIG waveguide, which then propagate along the thermal gradient towards a colder (a) and a hotter (b) region. (c) BLS measurements of the spin-wave reflection in a thermal gradient. The spatial spin-wave intensity distribution is shown for uniform temperature (top) and increasing temperature (bottom). White (black) color indicates high (low) spin-wave intensity. (d) Calculated spin-wave dispersion relations for a position near the antenna (dark blue) and at a distance of 3.5 mm to the antenna (pale red) illustrating the underlying mechanism of the spin-wave reflection. (e) Phase-resolved BLS interference measurements revealing the wavelength reduction with an applied thermal gradient (bottom) in comparison to uniform temperature (top). The interference between light that is scattered inelastically from spin waves and reference light with constant reference phase is measured as a function of the lateral position on the waveguide. A high (low) interference signal is indicated by white (black) color. (f) Comparison between extracted values of the experimentally observed spin-wave wavelength (symbols) and calculated wavelength values based on the measured temperature along the stripe (lines) for uniform temperature (black) and an applied thermal gradient (pale red). (g) Calculated spin-wave dispersion relations at two different positions. The temperature induced frequency shift of the curves causes an increase in the spin-wave wave vector and hence a wavelength reduction.

The resulting interference measurements for a waveguide with an applied thermal gradient and a reference measurement with a constant temperature are depicted in Fig. 2e. The upper panel of Fig. 2e shows the results of spin waves propagating in a waveguide with homogeneous temperature. A constant spacing between the interference minima is observed indicating that the wavelength remains constant along the waveguide. In contrast, for spin waves propagating towards the colder region (lower panel of Fig. 2e) a continuous wavelength reduction is detected.

This behavior can be understood by taking into account the change in the saturation magnetization and hence the frequency shift of the dispersion relation. Figure 2g shows the calculated dispersion relations for a position close to the excitation antenna (solid line) and at the end of the YIG waveguide (dashed line). For a constant excitation frequency, a continuous reduction of the spin-wave wavelength with temperature is obtained. A comparison of the calculated wavelength values with the experimental data (Fig. 2f) is in good agreement. For increasing distance from the antenna the experimental results exhibit an increasing deviation to smaller wavelength values even for the case of a constant sample temperature. Since this wavelength reduction appears in both measurements, it might be caused by a self-induced effect of the spin waves.

Another experiment of spin waves traveling towards higher temperatures has been realized in the configuration shown by the thermal image in Fig. 2b. Figure 2c shows BLS measurements of the spin-wave intensity distribution in the YIG waveguide. The microwave excitation frequency of 7.132 GHz was close to the FMR frequency of the waveguide and the applied temperature gradient in the waveguide was high enough to compensate the intrinsic wavelength transformation, which now acts opposite to the externally imposed wave vector modification. In a reference measurement with uniform waveguide temperature (upper panel of Fig. 2c), the spin-wave intensity distribution resembles a snake-like pattern which is understood to be the superposition of several transversal waveguide modes. The present case can be attributed to an interference of the two modes $n = 1$ and $n = 3$, where $n$ denotes the spin-wave mode number.

Measurements of the spin-wave intensity distribution in a waveguide with a temperature difference of $\Delta T = 16$ K (lower panel of Fig. 2c) show a drastically reduced spin-wave intensity in the regions far from the antenna (beyond dashed line at $x = 3.5$ mm) while there is an increase in the detected signal close to the excitation region (up to dashed line at $x = 1.5$ mm), both indicating a reflection of the spin waves in the thermal gradient.

The reflection can be understood in a similar fashion as discussed above. Increasing the temperature causes a downward shift of the dispersion towards lower frequencies. On propagating to a larger distance from the antenna, the temperature difference will induce a dispersion shift below the excitation line (7.132 GHz). Thus, spin waves with this given frequency can no longer exist and have to be reflected. The temperature gradient causes the formation of a "forbidden" region for these spin waves. Since the dispersion relations of the two waveguide modes $n = 1$ (solid lines) and $n = 3$ (dashed lines) are displaced in frequency, their critical points where reflection occurs do not coincide.

### 2.3. Control of anisotropic SW propagation by magnetization gradients

Spin waves in in-plane magnetized films are of interest for the realization of magnonic circuits. However, the dispersion relations of the corresponding spin waves, namely backward volume (BVMSWs) and surface spin-waves (MSSWs) are strongly anisotropic and do not coexist at the same frequency. BVMSWs propagate along the static magnetisation and MSSWs propagate perpendicularly, respectively. For a fixed saturation magnetisation and a fixed external magnetic field, the frequencies of MSSWs lay above the ferromagnetic resonance (FMR) frequency whereas the frequencies of BVMSWs lay below. A change in the propagation direction requires mode conversion between MSSWs and BVMSWs, which energy and momentum conservation obviously prohibit.

To overcome this problem we used an heat-induced approach to change the spin-wave propagation using graded index spin-wave media extending over sizes $\delta x$ much larger than the spin-wave wavelength $\lambda = 2\pi/|k| = \delta x$. A continuous transformation of spin-wave momentum in magnetisation gradients leads to highly-efficient mode conversion over a wide frequency range. We proposed and demonstrated experimentally as well as numerically mode conversion between backward volume and magnetostatic surface spin-wave modes using gradients of the saturation magnetization [31]. The magnetization landscapes are achieved via reconfigurable laser-induced thermal patterns [32], in which the BVMSW and MSSW modes coexist at the same frequency. The gradients perpendicular to the spin-wave propagation direction allow for the 90-degree rotation of the spin-wave wavevector via breaking the translational symmetry of the waveguide. Shape and orientation of the gradient control the conversion efficiency (see Figure 3).

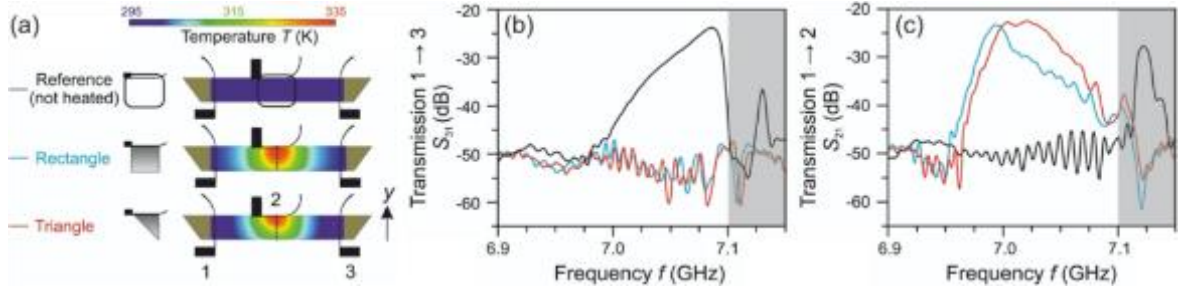

**Figure 3. Scheme of the experimental setup and transmission characteristics for different light intensity distributions [31].** The mode conversion process transforms BVMSW to MSSW modes. The external magnetic field $H_{ext}$ is oriented parallel to the waveguide. (a) Rectangular and triangular graded light intensity distributions are applied in the center between antennas 1 and 3 – at the position of antenna 2. The corresponding (rectangle: blue line, triangle: red line) SW transmission characteristics $S_{31}$ (b) and $S_{21}$ (c) are measured and compared with the reference measurement without heating (black line).

In order to obtain deeper insight into the conversion mechanism, numerical micromagnetic calculations were performed for different magnetization variations. Both the spin-wave phase structure and the distribution of the energy flow are revealed (see Figure 4) showing a highly anisotropic two-step spin-wave conversion process, which is in direct agreement with our experimental results. As opposite to the spin-wave momentum, the spin-wave energy is conserved due to the intersection of frequency bands of both spin-wave modes allowed in a non-uniform medium. Since the proposed conversion is a smooth process (the characteristic length scales of the gradients are much larger than the SW wavelength), it allows for the minimization of undesirable reflections and ensures high efficiency of the conversion in a wide range of spin-wave frequencies. Moreover, it is shown that the conversion mechanism can be further enhanced by exploiting the refraction at the edge of the magnetization gradient. Thus, the mode conversion using the triangular-shaped area of the magnetization gradient has shown higher efficiency in a wider frequency range than the rectangular one. The magnetization gradient-based SW guiding mechanism in in-plane magnetized films can be extended to the nanoscale using, e.g., ion implantation. This technique opens a way to magnonic circuits with in plane steering of the highly anisotropic SW modes. Moreover, the proposed approach of gradually changing material properties can be extended to waves in strongly anisotropic media in general.

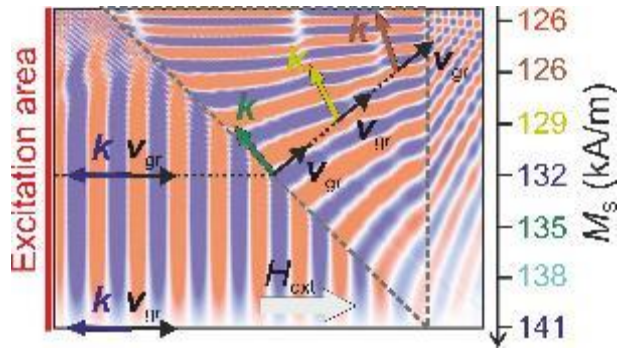

**Figure 4. Micromagnetic simulations of the spin-wave transformation [31].** The simulations are performed for a triangular gradient shape of the saturation magnetization $M_S$ at SW frequency $f = 7.0$ GHz. The spin wave is excited at the left of the picture and propagates to the right into the magnetization gradient region (dashed triangle) where the conversion takes place. The dynamic magnetization $m_z$ is color-coded (red: max, blue: min). A two-step conversion process is visible: first, the spin waves are refracted at the interface. Next, the transformation inside the gradient takes place.

### 2.4. Heat-induced magnonic crystal

Reconfigurable magnonic crystals, whose properties can be changed on demand (see reviews [37, 38]), attract special attention since they allow for tuning of the functionality of a magnetic element: the same element can be used in applications as a magnon conduit, a logic gate, or a data buffering element. An important step toward the realization and study of reconfigurable magnonic crystals was the demonstration that any two-dimensional magnetization pattern in a magnetic film can be created in a reconfigurable fashion by laser-induced heating [32]. By using a laser, a thermal landscape in a magnetic medium was created (see Figure 5) that resulted in an equivalent landscape of the saturation magnetization [32] and, hence, in the control of the spatial spin-wave characteristics. The setup used for the realization of the light pat- terns consisted of a continuous-wave laser as a light source, an acousto-optical modulator for temporal intensity control, and a spatial light modulator for spatial intensity control.

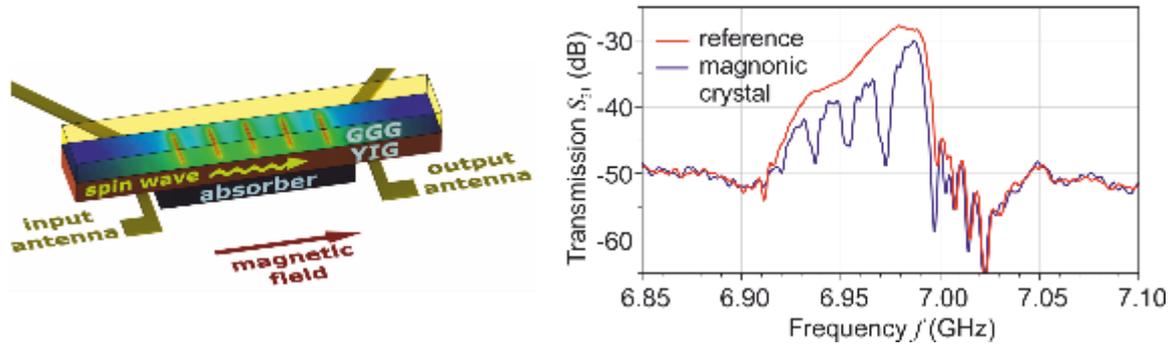

**Figure 5. Heat- induced reconfigurable magnonic crystals [32].** Schematic of the reconfigurable magnonic crystalwith a GGG/YIG/absorber multilayer system. A typical thermal landscape is shown on the waveguide (color code: red, maximum temperature; blue, minimum temperature). Right panel: Measured spin-wave transmission characteristic (solid blue line) in the thermal landscape (magnonic crystal with five periods, lattice constant 740 μm) and reference data (red) without the projected pattern. Formation of the band gaps in spin-wave spectrum are clearly visible.

In order to study the influence of the thermal gradient induced by the intensity patterns on the spin waves, we used a ferrimagnetic, 5 μm thick YIG waveguide grown on a 500 μm thick GGG substrate (see Figure 5). GGG is almost transparent, while YIG absorbed about 40% of the green light used in the experiment. In order to increase the efficiency of the heating, we used a black absorber coated on top of the YIG film. The proposed concept of a reconfigurable magnetic material was demonstrated and tested on the example of one- and two-dimensional magnonic crystals. The formation of band gaps in the spin-wave transmission characteristic is shown in the right panel of Figure 5. It was demonstrated that the positions of the band gaps can be tuned by the lattice constant of the magnonic crystal, while the widths of the band gaps were controlled by the laser light intensity [31]. This approach shown how heat can be used for the formation of any reconfigurable pattern for the control of spin-wave propagation.

### 2.5. Magnon-induced heat conveyer effect

Thermal gradients appear as a natural result of the thermalization of externally excited magnons. It is obvious that some part of the magnon energy thermalizes inside the magnon spectrum while another part is converted into lattice vibrations. We have measured the temperature gradient created by externally excited coherent spin waves using infrared thermography.

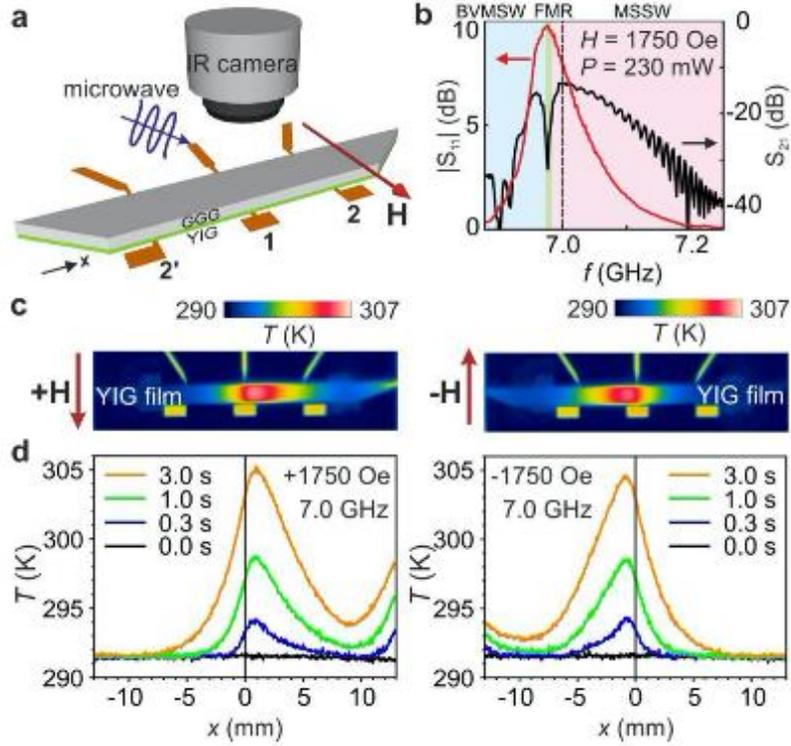

**Fig. 6. Observation of heat generation in a thin YIG film [33].** (a) Sketch of the setup for temperature distribution measurements in a YIG film on a GGG substrate using an infrared camera. The YIG sample is set on a distance of 100 μm away from the micro- wave input (1) and the output (2 and 2') antennas to prevent direct thermal contact. (b) Microwave absorption ($S_{11}$) and spin- wave transmission ($S_{21}$) spectra of the YIG film measured at 230 mW microwave power. Regions of magnetostatic spin waves are indicated by blue (BVMSW), green (FMR), and pink (MSSW) colors. The spin-wave transmission is maximal at a microwave frequency of 7.0 GHz (verti- cal dashed line). (c) Temperature distribution images of excited surface spin waves (7.0 GHz) for opposite orientations of the applied magnetic field H = ±1750 Oe. (d) Evolution in time of the temperature distribution profile of the YIG/GGG sample for different polarities of magnetic field.

For this study nonreciprocal surface magnetostatic spin waves propagating perpendicularly to the magnetization direction in a tangentially magnetized YIG film were used [33]. Figure 6a shows a sketch of the experimental setup. The experiments were performed using a single-crystalline YIG film of 30 μm thickness. The sample was placed close to a narrow microstrip antenna of 50 μm width. The distance between the YIG surface and the excitation antenna was kept 100 μm in order to increase thermal isolation of the sample. The YIG film was magnetized by an in-plane magnetic field of $H$ = 1750 Oe. Figure

6b shows microwave absorption ($S_{11}$) and spin-wave transmission ($S_{21}$) spectra of the YIG film.

In the presence of travelling spin waves excited by a continuous microwave signal at 7 GHz frequency, a heating of the sample was observed around the input microstrip antenna. The thermal distribution along the YIG sample is shown in Fig. 6c. In the absence of a magnetic field, the temperature of the sample remains equal to the room temperature when the microwave is switched on. Once a sufficiently large field is applied, the temperature of the sample increases and results in the thermal distribution shown in Fig. 6c. Maximal heating was expected at the excitation point, i.e. directly over the microstrip antenna. However, an unsymmetrical distribution with the shift of the temperature maximum away from the excitation antenna of up to a few millimeters was observed in this case. We have found a correspondence between this shift and the group velocity of the surface spin waves. The asymmetry of this distribution was understood as a result of the interplay between the unidirectional heating by MSSWs and the diffusion of heat into the colder part of the YIG film. It is in agreement with the nonreciprocal excitation of surface waves. In contrast, reciprocal backward volume magnetostatic spin waves demonstrate a symmetrical heat distribution with the maximal temperature at the position of the excitation antenna.

Thus, we can conclude that a magnetically controllable heat flow caused by a spin-wave current in a magnetic insulator was detected for surface magnetostatic spin waves [33]. The direction of the flow can be switched by changing the direction of the applied magnetic field. This phenomenon was understood as unidirectional energy transfer by the excitation of nonreciprocal spin waves.

For a spin wave to convey energy efficiently, the *k* and *-k* populations of spin-wave states should be different. Here, *k* represents the wave vector or the momentum. A simple method of realizing this difference is the use of nonreciprocal magnetostatic surface spin waves of ferromagnetic materials.

To realize this situation, we have used the same setup schematically shown in Fig. 6a. We have observed that nonreciprocal spin waves in the range of microwave frequencies can convey energy and emit heat at a far distance (the end of the YIG sample in Fig. 6c and Fig. 6d) to create a negative temperature gradient different from conventional microwave heating. The unidirectional heat transfer effect observed in this study can be explained by the spin-wave energy, which is proportional to the spin-wave particle (magnon) number. In the present system, the energy current is carried by the MSSW spin-wave current. There are two possible mechanisms for spin waves to release heat at the end of the sample (Fig. 6c). One is the suppression of the reflection of the spin waves from the bottom surface MSSWs to top surface MSSWs. This mechanism is the most effective one for relatively thick samples. However, in the thin single-crystalline YIG film used in our experiments, where the MSSW interaction of the two surfaces is strong, effective MSSW reflection can be suppressed as a result of spin-wave multi-reflections by using a sharp 30° angle for the spin-wave waveguide. The other mechanism is the occurrence of two-magnon scattering processes at the edge of the sample where the effective fields are not uniform owing to a magnetic inhomogeneity with respect to magnetization and magnetic anisotropy. As a result, both mechanisms pump all the energy of spin waves into the lattice as heat, and the heat conveyer effect exhibited there was measured by the lattice temperature using an infrared camera.

## 3. Spin caloritronics and magnon Seebeck effect

### 3.1. Phonon contribution to spin pumping and transversal SSE phenomena

Although there was only minor interest in spin caloritronic effects over a long time, during the last decade, they have come to be recognized as a very promising route to the development of future spintronic devices. Two factors have played a particularly important role in the spin caloritronic revival: (1) recent pioneering results in the field of spintronics and (2) the constantly growing problem of parasitic heating at hot spots in CMOS devices. One of the foremost aims in spintronics is the simultaneous application of electron spin and charge for information transfer and processing to improve the ratio of processed data versus parasitic heating. A very interesting question therefore, is whether parasitic heating can be turned into a helpful phenomenon.

In this context, an important milestone was the discovery of the spin Seebeck effect (SSE) by the group of E. Saitoh [41]. It was shown that in a magnetic metal subject to a thermal gradient a spin current is generated. The authors have postulated that the effect is based on different temperature dependencies of the scattering rates for counter-propagating electrons having opposite spins. SSE has been shown experimentally using a thin-film Permalloy (Py)/ Pt structure. In the Pt layer the inverse spin Hall effect (ISHE) [42] is used to detect a spin current flowing from the Py to Pt layer. Afterwards, the same group published equally groundbreaking experimental evidence of the existence of the SSE in a magnetic insulator [43]. They used La doped YIG with a Pt layer which played the role of a spin current detector via the ISHE. During further investigations, SSE was detected in many different magnetic insulators [44,45]. These experiments demonstrated the dominant role of magnons in the SSE and triggered detailed investigation of the connection between the magnon Seebeck effect with the basic properties of magnon and phonon interactions.

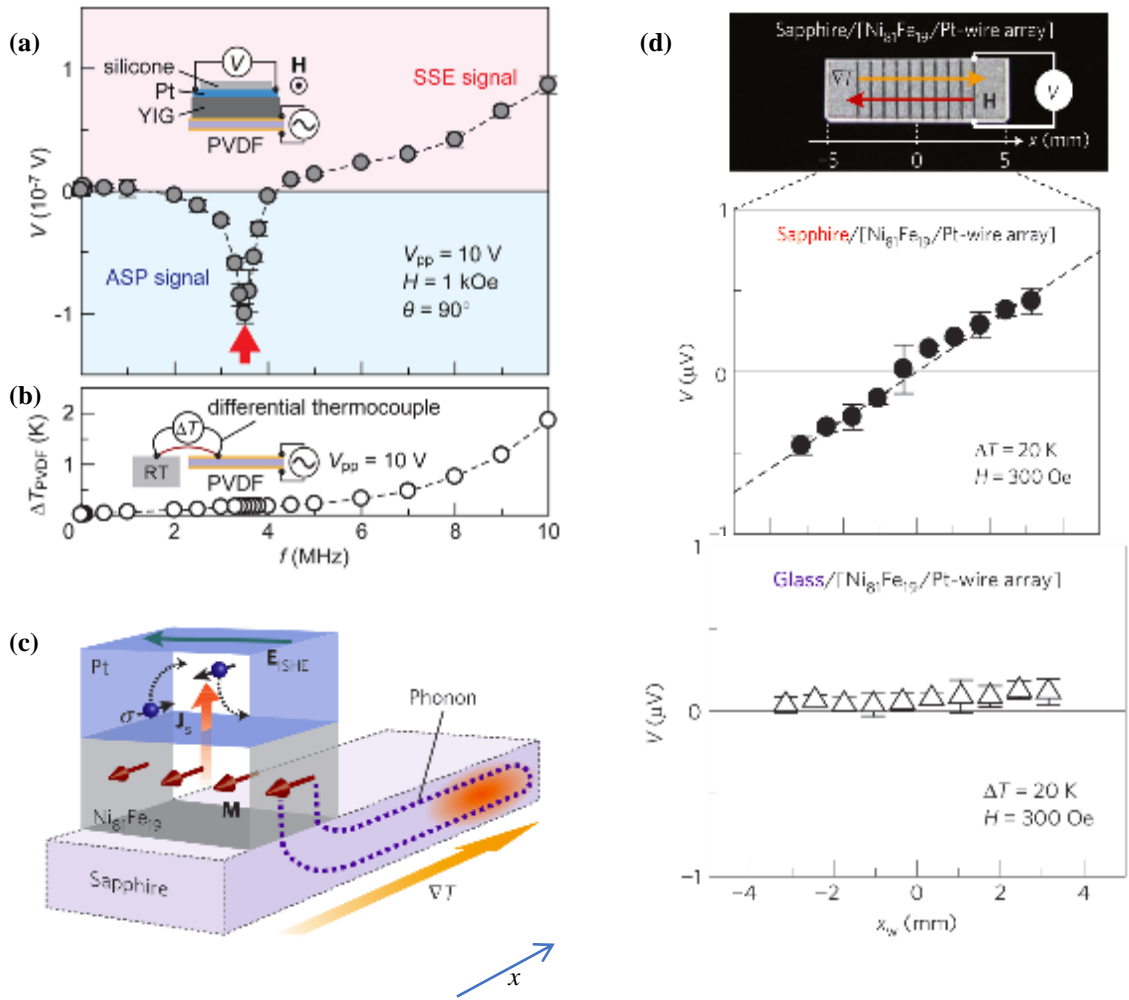

**Fig. 7. Acoustic spin pumping and long-range spin Seebeck effect [46,47].** Frequency dependence of ISHE voltage measured between the ends of the Pt layer in the Pt/YIG sample (a) and temperature of the attached polyvinylidene-fluoride (PVDF) film based piezoelectric acoustic actuator (b). (c) Scheme of the phonon mediated SSE and ISHE in the sapphire/[Py-Pt-strip] sample. (d) SSE voltage at the Pt/Py-strip array sample. The interval between the adjacent Pt/Py wires is 0.7 mm. No voltage is detected in the case of a glass substrate with the short-distance phonon free-path.

The possible contribution of the phonon flows in the formation of the spin Seebeck effect was firstly reported in the work, where the ability of acoustically mediated spin pumping by low-energy coherent phonons propagating across [46] and along [47] a Pt/YIG bilayer was discovered: The sound wave generated by a piezoelectric actuator modulates the distribution function of magnons in the YIG layer and results in a pure spin-current injection into the Pt layer across the Pt/YIG interface (see the negative ISHE voltage peak in Fig. 7a associated with the resonance excitation of the sound wave). The phenomenon was separated from the longitudinal SSE which also results in a spin current across the Pt/YIG layer interface: As opposed to the case of the transversal SSE, this spin current propagates along the thermal gradient created by the hot acoustic transducer (please note that Fig. 7b shows no peculiarity in the transducer temperature at the ASP signal frequency).

Even more importantly, the crucial role of a phonon-mediated spin pumping in the transverse SSE has been revealed in Ref. [47]: The transverse SSE voltage was detected in the array of magnetically and electrically separated Permalloy/Pt wires (Fig. 7c-d). Moreover, the voltage dependence shown in Fig. 7d is detected only if the Pt/Permalloy structure is grown on a single- crystal sapphire substrate characterized by a long free-path of thermal phonons at frequencies around 20 THz. No voltage is detected in the case of a glass substrate with much shorter phonon free-path (Fig. 7e). Obtained results about the important role of the phonon transport in spin-caloric phenomena perfectly correlate with results on a phonon driven spin distribution in GaMnAs samples [48] as well as with the discovery of the giant spin Seebeck effect, which is mediated by direct phonon–electron drag in a non-magnetic semiconductor (InSb) [49].

Interesting results have been reported recently in Ref. [50] in which nonlinear relaxation between magnons and phonons in an insulating ferromagnet is investigated theoretically. Therein, magnons and phonons are described by equilibrium Bose-Einstein distributions with different temperatures and the nonlinear heat current from magnons to phonons is calculated microscopically in terms of the Cherenkov radiation of phonons by magnons. In subsequent studies, the heat transfer through an insulating ferromagnet sandwiched between two insulators has been investigated theoretically [51]. Depending on the geometrical sizes of the heterostructure and the mean free path of phonons generated by magnons, two qualitatively different regimes in the nonlinear heat transport through the interfaces have been revealed. In the framework of the kinetic approach based on the Boltzmann equation for the phonon distribution function exploited in Ref. [52], the phonon heat transfer in the heterostructure has been described [52]. The dependence of the ISHE voltage on the bath temperature and on the thickness of the layers has been compared with experimental data. Very recently, temperature dependence of the magnon-phonon energy relaxation time in a ferromagnetic insulator has been analyzed theoretically [53]. When the thickness of the ferromagnetic layer is much greater than the phonon-magnon scattering length, the magnon temperature dependence on the frequency has been predicted to exhibit features related to specific characteristic times of the system.

## 3.2. Heat induced damping modification and magnon relaxation as longitudinal SSE source

At the first stage, we have shown the longitudinal SSE effect in an electromagnetically driven Pt/YIG bilayer [54]. The ISHE voltage $V_{ISHE}$ was measured in a 10 nm thick 3×3 mm$^2$ Pt square deposited onto a 2.1 μm thick single-crystal YIG film. The spin currents across the Pt/YIG interface were induced by parametrically excited magnons in the YIG film as well as by thermal electron density fluctuations in the microwave heated platinum. In the first process spins are injected into the Pt layer, while in the second one (longitudinal SSE) spins are ejected from the Pt layer into the YIG layer. Thus, these two effects show different polarities in $V_{ISHE}$. The contributions of these processes were separated by the suppression of parametric magnon injection due to a proper choice of the bias magnetic field as well as by the observation of their temporal dynamics. In addition, this result in combinations with the results reported in Refs. [55-58] has proven that all magnons independently on their propagating or standing nature and on the wavelength efficiently contribute to the spin pumping and can be measured in a form of ISHE voltage.

This successive experiment [57] focused on the interplay between temperature gradient dependent phenomena and coherent spin-wave dynamics. The temperature difference $\Delta T_{Peltier}$ across a YIG/Gadolinium Gallium Garnet (GGG)/YIG/Pt hetero-structure was created by a Peltier element mounted on top of the Pt layer. In order to enhance the temperature flow from the sample, one of the YIG surfaces was covered with a sapphire substrate that was connected to a heat bath (sapphire is a good thermal conductor). The magnetization precession was excited by a microstrip antenna that was placed above the Pt layer with an intervening isolation layer (see Fig. 8a). The temperature difference was monitored using an infrared camera.

The applied temperature difference $\Delta T_{Peltier}$ leads to the appearance of the longitudinal SSE: an imbalance between the magnon and electron temperatures causes spin transfer across the Pt/YIG interface. As a result, a spin transfer torque (STT) acts on the magnetic moments on the YIG film surface, and their precessing motion can either be enhanced or suppressed depending on the sign of the temperature difference and, thus, the direction of the spin transfer (Fig. 8b). This change in the damping is equivalent to a variation of the ferromagnetic resonance (FMR) linewidth $\Delta H$ that can be detected by measuring the DC ISHE voltage UISHE caused by spin pumping into the adjacent Pt layer (Fig. 8c).

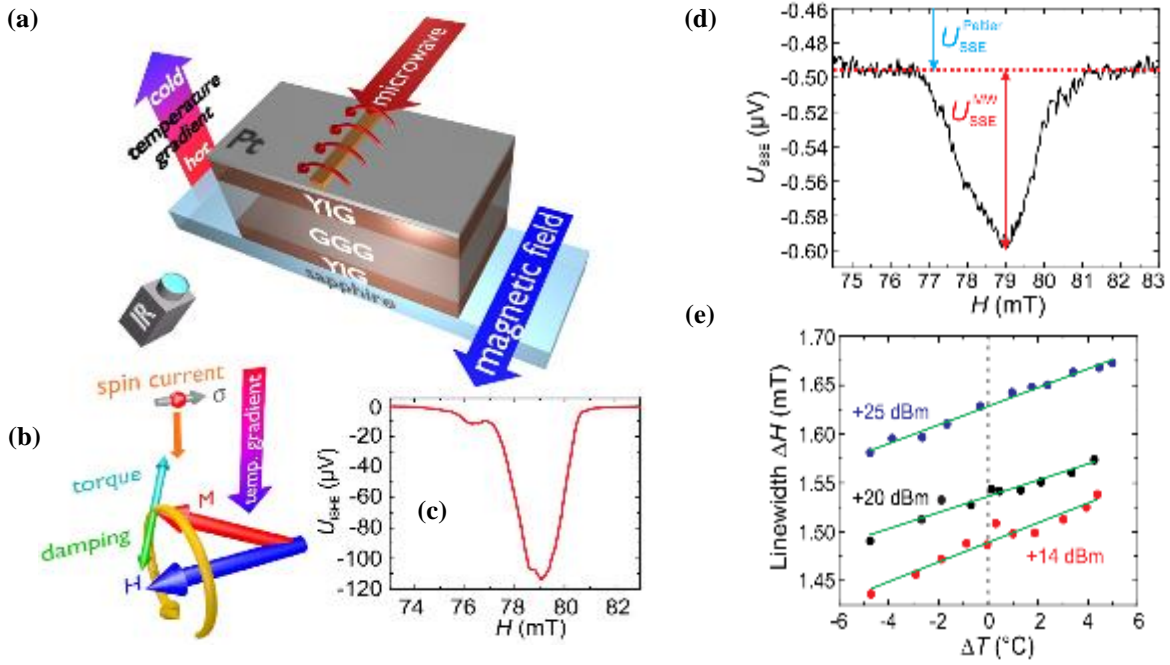

**Fig. 8. Heat-induced damping modification in YIG/Pt hetero-structures [57].** (a) Experimental setup. (b) Possible mechanism for the heat-induced damping variation in the YIG film. The red sphere represents a spin polarized electron and the arrow through the sphere indicates the spin polarization vector **σ**, which is aligned anti-parallel to the direction of the bias magnetic field **H**. The magnetization **M** precesses around **H**. (c) Typical example for a measurement of $U_{ISHE}$ as a function of $H$. (d) Spin Seebeck voltage $U_{SSE}$ composed of $U_{Peltier}$ generated by $\Delta T_{Peltier}$ and $U_{MW}$ created by $\Delta T_{MW}$. (e) Resonance linewidth $\Delta H$ as a function of the temperature difference $\Delta T$ for different microwave powers.

At the same time, the measured $U_{ISHE}$ contains a spurious thermal component $U_{SSE}$ induced by the externally applied $\Delta T_{Peltier}$ and by an additional temperature difference $\Delta T_{MW}$ that is created due to the internal YIG heating by decaying spin waves (Fig. 8d). We have introduced a method to distinguish between the spin pumping by coherent magnons and the LSSE contribution from the thermal magnon bath. It has been found that the internally induced $\Delta T_{MW}$ leads to an increase of the ISHE voltage of around 0.1%, and thus can be neglected in the majority of the spin pumping experiments with low-damping magnetic materials. It is worth to note that this value correlates with our early data on the contribution of the relaxed magnons to the spin-pumping temporal dynamics [58].

The linewidth $\Delta H$ as a function of the temperature difference $\Delta T_{Peltier}$ is shown in Fig. 8b for different microwave powers. It is clearly visible that $\Delta H$ decreases for one polarity of $\Delta T$ and increases for the other. For a temperature difference of $\Delta T \cong 4°C$, the linewidth changes about 6%, independent of the microwave power. This independency is expected since the thermally generated spin currents do not depend on the applied microwave power. The spin-wave damping, i.e., the linewidth at $\Delta T = 0°C$ is larger for higher microwave powers, which is attributed to the onset of non- linear effects.

Assuming that the observed heat-induced damping variation is due to an effective spin current representing thermally induced torque, we have estimated the spin transfer due to the temperature difference across the Pt/YIG interface. It turns out that the heat-induced spin current density per 1°C is of the order 10 A/m . It should be emphasized that this value is one to two orders of magnitude higher than those reached in Ref. [59] using the spin Hall effect induced scattering of a charge current flowing through the Pt layer. Taking into account the large magnitude of the heat induced STT in the comparison with the relatively small temperature difference across the Pt/YIG interface (less than 1°C), we might consider that bulk effects connected with a long-distance magnon and phonon transport could contribute to the observed STT voltage along with the interfacial interactions.

### 3.4. Time-dependent magnon Seebeck effect

In order to clarify the role of the bulk effects in formation of the longitudinal magnon Seebeck effect we have performed time-resolved measurements of the LSSE voltage $U_{SSE}$ in Pt/YIG bilayers heated by laser pulses [60] and microwave pulses [61]. The design of the laser-heating experimental setup is shown in Fig. 9a. The Pt strip was simultaneously utilized as a fast-response resistive sensor to measure the temperature at the surface of the YIG film.

Figure 9b shows the profile of the measured SSE voltage signal (triangles) in comparison to the variation in the temperature of the Pt strip (see inset) and with the laser pulse front profile (dashed line). A drastic difference in the voltage and the temperature dynamics is visible: The SSE signal rise time of 340 ns, obtained by an exponential fit to the data, is nearly four orders of magnitude shorter than the temperature rise time of 2 ms. Both dynamics are independent of the laser power, and thus are free from spurious nonlinear phenomena. Thus, it can be concluded that the temperature of the system has no direct correlation with the fast timescale of the SSE.

In order to understand the fast rising of the SSE, we proposed a model where the spin Seebeck voltage consists of two contributions: an interface effect and a bulk contribution from the magnon motions. It might be expressed as $U_{SSE} \propto \alpha (T_N - T_M) + \beta \int_y \nabla T_y \exp(-y/L) \, dy$, where $T_N$ is the phonon temperature (equal to the electron temperature) in the normal metal, $T_M$ the magnon temperature at the interface, $\nabla T_y$ the phonon thermal gradient perpendicular to the interface, and $L$ the effective magnon diffusion length. The parameter α defines the coupling between the electron bath in the normal metal and the magnon bath in the magnetic material. The coupling parameter β specifies the magnon-magnon coupling within the magnetic material. The integration is performed along the phonon thermal gradient over the thickness of the magnetic layer.

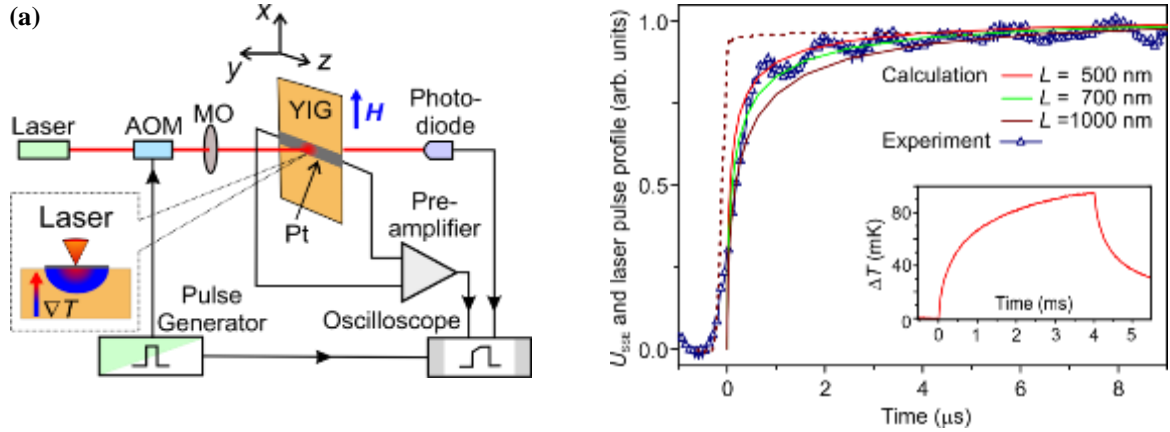

**Fig. 9. Sub-microsecond fast temporal evolution of the longitudinal magnon Seebeck effect [60].** (a) A continuous laser beam (wavelength 655 nm), modulated by an acousto-optical modulator (AOM), is focused down on a 10 nm thick Pt strip, deposited on a 6.7 μm thick single-crystal YIG film, by a microscope objective (MO). The laser intensity profile is monitored by an ultrafast photo-diode. An in-plane magnetic field $H = \pm 200$ Oe was applied to the YIG film. The heated Pt creates a thermal gradient perpendicular to the Pt/YIG interface (see inset). (b) Time profiles of the laser intensity (dashed line), and the SSE voltage ($U_{SSE}$) at a laser power of 130 mW. The maximal value of the $U_{SSE}$ is 2 μV. $U_{SSE}$ changes its polarity by reversing the direction of the magnetic field. Solid lines represent numerical calculations of $U_{SSE}$ for different effective magnon diffusion lengths $L$. The inset shows the time profile of the variation of temperature in the Pt strip upon heating with a 4 ms long laser pulse.

In order to determine the phonon thermal gradient $\nabla T_y$, we numerically solved the 2D phonon heat conduction equation for the Pt/YIG bilayer using the COMSOL Multiphysics simulation package. First of all, we simulated the temporal evolution of the average temperature in Pt which is as slow as it was observed in the Pt-resistance measurement ($\cong$ 3 ms). It is explained by the large heat capacity and the volume of the system. On the other hand, the thermal gradient shows a much faster rise time from about 50 ns at the Pt/YIG interface to 5 µs at a position 1 µm apart from the interface. Furthermore, the simulations show that the lateral heat flow in the YIG film and the heat transport within the Pt strip have minor influences on the average thermal gradients.

The pronounced difference between the rise time of the LSSE voltage and the fast rise of the thermal gradient at the interface of the Pt/YIG bilayer indicates once again that the timescale of the SSE cannot be explained only by the interfacial mechanisms and must be influenced by a rather slower process. On the basis of this argument, the first term of the $U_{SSE}$ equation, which is proportional to the phonon thermal gradient directly at the interface, can be considered as static over the timescale of interest (>50 ns). Using the phonon thermal gradient data, obtained from the COMSOL simulations, we calculated the integral term of $U_{SSE}$ for different magnon propagation lengths as a function of time (see Fig. 9b). Clearly, our model matches the experimentally observed time scales of the SSE. For a comparison of the calculated voltage USSE with the experimental results we estimate the effective magnon diffusion length – the depth inside the YIG layer over which the thermal gradient is crucial for the SSE. It is about 700 nm and this value agrees with theoretical calculations [62, 63]. Thus, we can conclude that thermal magnons in a depth of up to a few hundred nanometers in the YIG film effectively contribute to the SSE. It is notable that this finding rules out the possibility of parasitic interface effects like, for example, the anomalous Nernst effect in a magnetically polarized Pt layer—which could potentially distort the measurements.

### 3.5. Direct measurements of magnon temperature in thermal gradients

In order to understand magnon currents generated by a thermal gradient in a magnetic insulator great efforts have been undertaken for the direct probing by BLS of thermal magnon currents in pure YIG single-crystal films, which have the smallest magnetic damping among all known practical magnetic materials (not involving any intermediate entities like inverse spin Hall and spin Seebeck effects). The spatial distribution of the magnon density created by a linear temperature gradient in in-plane magnetized YIG films has been explored using space- resolved BLS spectroscopy with a spatial resolution of 100 µm. Space-resolved infrared thermometry was used for a simultaneous two-dimensional mapping of the phonon thermal gradients in the film plane. Two main experimental approaches were used: 1) Spatial mapping of the integral magnon density, and 2) BLS measurements of the thermally dependent energy-momentum distribution of dipolar-exchange low-energy magnons which have wavenumbers below the maximal wave vector of $2.5 \times 10^5$ cm$^{-1}$ accessible by wave-vector resolved BLS [64]. Since each thermal exchange magnon reduces the total magnetization of a magnetic system by one Bohr magneton, the local magnetization is a measure of the local magnon population, and hence the magnon temperature $T_m$. It follows that spatial variations in the magnon temperature of a magnetic system can be determined through spatially resolved measurements of its magnetization $M$. However, the changes in magnetization that must be detected are very small, making its measurement a challenging task. However, BLS spectroscopy offers an elegant solution. Our technique allows for a measurement of the magnetization via the detection of the frequency shift of a specific thermal exchange magnon mode (Fig. 10a,b).

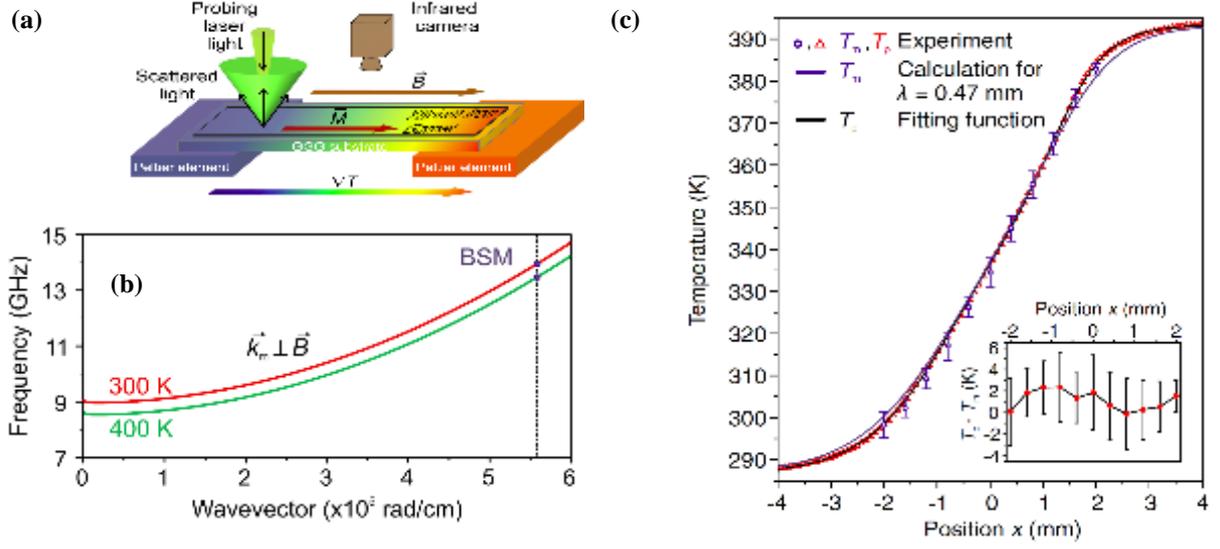

**Fig. 10. Thermal excitation and BLS probing of a lateral magnon current [65].** (a) Two opposite ends of a YIG film (on a GGG substrate) are placed on Peltier elements to create a lateral thermal gradient $\Delta T$. The film is magnetized in plane with a magnetic field $B = 250$ mT applied parallel to $\Delta T$. An IR camera is used to obtain thermal images of the system. (b) Dispersion relations for magnons propagating perpendicular to the magnetization at temperatures of 300 K and 400 K in a 6.7 µm thick YIG film. Bullets show the positions of the back scattering magnon (BSM) mode whose frequency shift indicates the temperature dependent change of the saturation magnetization of the YIG film. c) Measured phonon $T_p$ (triangles) and magnon $T_m$ (circles) temperatures along the sample. The $T_p$ data are fitted with a Boltzmann sigmoid function. The profile of $T_m(x)$ is numerically calculated for the characteristic length parameter $\lambda = 0.47$mm. The inset shows the difference between $T_p$ and $T_m$ with a 95% confidence level.

On the one hand, due to its small wavelength ($k=5.67\times10^5$ cm$^{-1}$) this mode is localized on a sub-micrometer length scale. On the other hand, the wavenumber of this mode is strictly defined by the momentum and energy conservation laws in the process of inelastic scattering of the probing laser light [65]. Thus, a thermally dependent frequency shift of the magnon spectrum results in exactly the same shift in frequency of the scattered light (Fig. 10b). This frequency shift corresponds unambiguously to the change of the saturation magnetization which can be calibrated by the magnon temperature change by probing a uniformly heated sample with $T_m=T_p$.

In an externally created temperature gradient $\nabla T$ (Fig.10a), the magnon and the phonon temperatures $T_m$ and $T_p$ depend on thermally induced phonon and magnon transport and are coupled through the magnon-phonon interaction. In Fig. 10c the measured phonon and the magnon temperatures are plotted as a function of the position along the YIG film. It is evident that Tm follows the trend of $T_p$ within the limit of experimental uncertainty. The difference between $T_p$ and $T_m$ shown in the inset does not exceed 2.8% of $T_p$.

Within the framework of a one-dimensional model as described in Ref. [66] these temperatures obey the equation $\frac{dT_m^2(x)}{dx^2}+\frac{(T_p(x)-T_m(x))}{\lambda^2}=0$, where $\lambda$ is a characteristic length scale proportional to the square root of the magnon-phonon relaxation time. Theoretical estimations of $\lambda$ in YIG have varied significantly from 0.85 to 8.5 mm [59]. From our experimental data on $T_m(x)$ and $T_p(x)$, we calculated the maximal possible value $\lambda_{max}=0.47$ mm. This value is roughly one order of magnitude smaller than the one estimated by Xiao et al. [62] for YIG using the experimental spin Seebeck data of Ref. [41].

## 4. Role of phonon modes in the formation of Bose-Einstein condensate of magnons

### 4.1. Magnon Bose-Einstein condensation

Bose–Einstein condensation [67-69] is one of the most exciting and intriguing quantum phenomena. Generally speaking, this process can be described as the coalescence of bosons (including real atoms and Cooper pairs of fermions, and quasiparticles of different nature) into the same quantum state and, thus, constituting a specific state of matter described by a single wave function – a Bose–Einstein condensate (BEC). The prerequisite of the BEC formation is exceeding the threshold density of the particles at the given temperature. This requirement justifies the necessity of cryogenic conditions for the observation of the BEC in a gas of real atoms [70]. Besides of the fundamental importance of the BEC-related phenomena such as superconductivity and superfluidity, they are thought to be at the frontline for the potential use in the next generation of electronic devices. A major challenge in this area is the realization of the room-temperature macroscopic quantum state computing.

The first theoretical predictions that a BEC can occur at higher (or even at ambient) temperatures were made by Fröhlich in 1968 [71]. On the way to ambient-temperature BECs, the approach is to use quasiparticles instead of real atoms. Quasiparticles are excitations of a many-body system, the vast majority of which are bosons and therefore can form a BEC. Compared to the real atoms, quasiparticles have many orders of magnitude smaller effective mass, which decreases their density threshold for the BEC observation. However, unlike real atoms, quasiparticles decay and decohere with time, making the experimental observation of their BEC very challenging. In order to compensate the natural decay, it is necessary to externally add an excess of quasiparticles. In such a case, the condensation is possible, if the flow rate of energy pumped into the system exceeds a critical value. Magnons, being the quanta of spin waves, are bosons. Thus, in the thermal equilibrium, they can be described by the Bose–Einstein statistics with the zero chemical potential and a temperature-dependent density. At cryogenic temperatures, the magnon BEC was observed in a superfluid $^3$He [72]. Theoretical foundations for the room-temperature magnon Bose–Einstein condensation were developed in 1989 by Kalafati and Safonov [73, 74]. These predictions resulted in the first experimental observation of a magnon BEC at room temperature in Yttrium Iron Garnet ($Y_3Fe_5O_{12}$, YIG) [13]. There were important preconditions for this observation. The first is the selection of the material. YIG films provide a very long spin-lattice relaxation time of up to 1 $\mu$s. Thus, the lifetime of magnons is sufficiently longer than the magnon-magnon thermalization time due to the two- and four-magnon scattering processes, which can be as low as 10–100 ns. As these scattering events conserve the number of magnons, a quasi-equilibrium state for the magnon gas can be realized with a nonzero chemical potential. The influence of three-magnon processes can be significantly reduced by a proper selection of the magnetic field and the magnon excitation frequency. The second precondition is the fast (on a time scale of several orders of magnitude shorter than the relaxation time) creation of an overpopulation of the magnon spectrum. It was realized with the use of the parametric pumping, which offers an efficient way to excite a large number of magnons in the low-frequency part of the magnon spectrum, which then can thermalize into magnons with the minimum frequency. Using this technique, a pumped magnon density of above $10^{19}$ cm$^{-3}$ can be reached [75, 76]. Although this density is much smaller than that of thermal magnons at room temperature ($\sim 10^{22}$ cm$^{-3}$), this increase is sufficient to trigger the BEC formation.

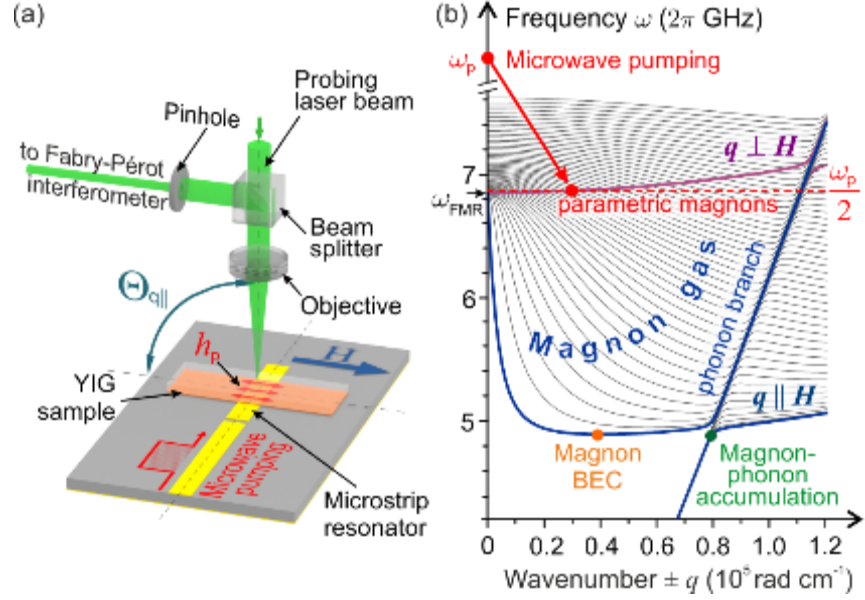

**Fig. 11. Schematic illustration of the BLS experimental setup** [77]. The resonator concentrates the applied microwave energy and induces a pumping microwave Oersted field $\mathbf{h}_p$ oriented along the bias magnetic field $\mathbf{H}$, thus realizing conditions for the parallel parametric pumping. The probing laser beam is focused onto a YIG film placed on the top of the microstrip resonator. The light inelastically scattered by magnons is redirected to a Fabry-Pérot interferometer for the frequency and intensity analysis. The wavenumber-selective probing of magnons with wavevectors $\mathbf{q} \parallel \mathbf{H}$ is realized by varying the incidence angle $\Theta_{q_\parallel}$ between the field $\mathbf{H}$ and the probing laser beam (a). The magnon-phonon spectrum of a 6.7 $\mu$m-thick YIG film is calculated for $H = 1735$ Oe. 47 thickness modes with $\mathbf{q} \parallel \mathbf{H}$ are shown. The upper thick curve shows the most effectively parametrically driven [78] lowest magnon mode with $\mathbf{q} \perp \mathbf{H}$. The arrow illustrates the magnon injection to the frequency $\omega_p / 2$ slightly above the ferromagnetic resonance frequency $\omega_{FMR}$ (b)

Another experimental challenge was to detect magnons in the BEC. The magnon spectrum of an in-plane magnetized YIG film is highly anisotropic in relation to the direction of magnon's wavevector. This anisotropy originates from the dipolar interaction and together with the isotropic influence of the exchange interaction leads to the appearance of the global energy minimum at the longitudinal branch of the magnon spectrum at a relatively high wavenumber $q = 4 \times 10^4 \,\text{rad}\,\text{cm}^{-1}$ (see Fig. 11b). This fact makes the conventional electromagnetic detection of the condensed magnons hard to be realized. However, the optical detection using BLS (via the inelastic scattering of photons on magnons) gives the necessary wide-range wavevector sensitivity together with decent frequency resolution. The BLS experiments were performed with time resolution, wherein the start of the pumping pulse triggers a stroboscopic clock for a counter of scattered photons. The main elements of the typical experimental setup for a BLS experiment are sketched in Fig. 11a.

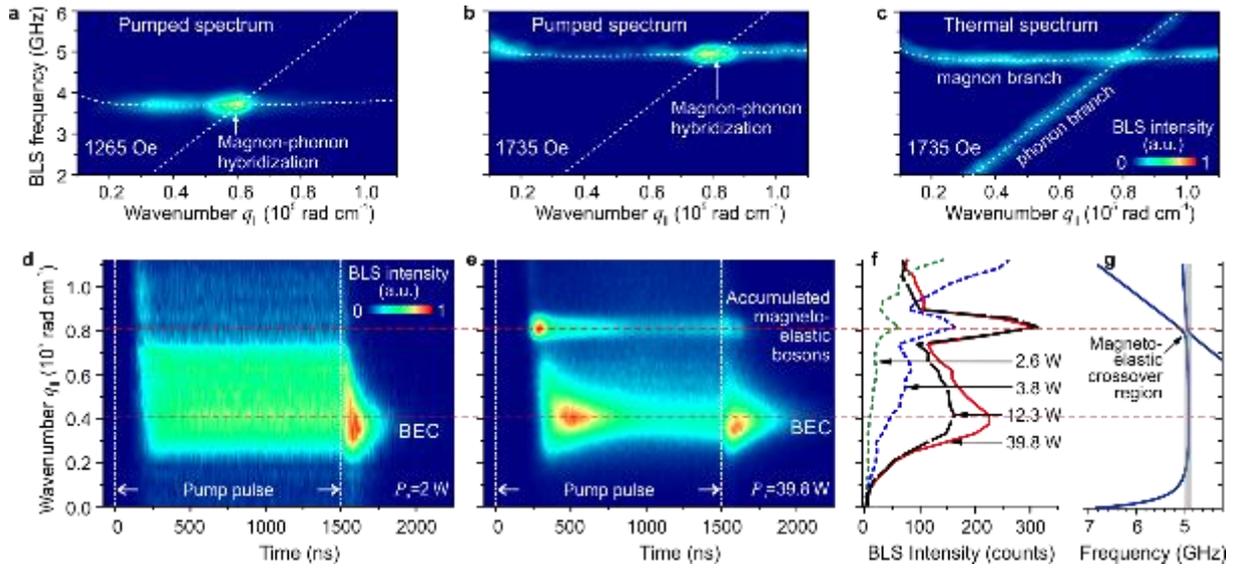

**Fig. 12. Magnon-phonon spectra and their population under different pumping conditions** [79]. White dashed lines represent the calculated dispersion relation for the lowest magnon branch ($q \parallel H$), hybridized with a transversal acoustic mode. Strong population of the magnon-phonon hybridization region is clearly visible in the parametrically pumped spectra (a-b) measured in the case of a 500 μm–wide pumping area. The width of the population peak is of the order of the wavenumber resolution of the experimental setup, i.e. $\pm 0.02 \times 10^5 \,\text{rad}\,\text{cm}^{-1}$. No distinct peculiarities in the thermal spectrum measured for zero pump power are detected (c). Temporal dynamics of the magnon gas density for narrow (d) and wide (e) pumping areas. No accumulation effect is visible in (d) due to a strong leakage of fast magneto-elastic bosons from the narrow pumping area of 50 μm width. Pronounced accumulation of magneto-elastic bosons is evident for the broad pumping area of 500 μm width (e). The panel (f) shows the pumping power dependencies of the quasiparticle density integrated by time. In all measurements BLS data are collected in the frequency band of 150MHz near the bottom of the magnon spectrum (see the shaded area in the panel (g)). The bias magnetic field is H = 1735Oe.

## 4.2. Condensation of mixed magnon-phonon states

Considering a magnon BEC and its excitations, the magnon-phonon interaction usually is taken into account only as a general decay mechanism of magnons to the lattice. However, the magnon-phonon scattering processes can significantly modify the scenario of thermalization of a pumped magnon gas. The magnon dispersion curvature in a vicinity of the hybridization between magnon and transverse phonon modes experiences strong modifications (see Fig. 11, b) [79]. This leads to a significant change in the magnon velocities as well as scattering efficiencies and appearance of magnon-phonon hybrid quasiparticles. The thermalization scenario also undergoes significant modifications near the crossing point, leading to the so-called bottleneck accumulation of hybrid magneto-elastic bosons [77]. It was found that the transfer of quasiparticles toward a BEC state is almost fully suppressed near the intersection point between the magnon and phonon spectral branches. Such a bottleneck leads to a strong spontaneous accumulation of the quasiparticles trapped near the semi-linear part of the magnon-phonon hybridization area.

The spectral characteristics of an overpopulated magnon-phonon gas were studied at room temperature in a YIG film by means of previously described BLS spectroscopy (see Fig. 11a). Following the approach, which was established in previous experiments on a magnon BEC [13,15,16], magnons were injected into the in-plane magnetized YIG film via parallel parametric pumping [80]. Surprisingly, another spectral point of quasiparticle's accumulation was found far away from the global energy minimum of a pure magnon spectrum (compare the orange and green dots in Fig. 11b). BLS intensity maps representing the population of the magnon-phonon spectrum are presented in Fig. 12a-c as a function of frequency and wavenumber $q_{\parallel}$ ($\mathbf{q} \parallel \mathbf{H}$) for two different bias magnetic fields and a relatively small pumping power $P_p$ of 2.6 W. In spite of the fact that the threshold for magnon BEC formation is still not reached at such power levels, one can see an intense population peak near the region of the hybridization between the magnon and the transversal acoustic phonon dispersion branches (see white lines in Fig. 12a-c). As the bias magnetic field is shifted from 1265Oe to 1735Oe, the population peak shifts in frequency and wavenumber together with the magneto-elastic crossover region. At the same time, there are no peculiarities in the thermal spectrum, measured at the same conditions, but without the application of pumping (see Fig. 12c).

It is important that the population peak is visible only if a sufficiently wide microstrip resonator is used to create the pumping microwave field $h_p$ (see Fig. 11a). This then confirms that the accumulation occurs into the magnon-phonon hybridization region, where quasiparticles possess rather high group velocities – the widening of the width of the microstrip reduces their leakage from the pumping area. Obviously, the accumulation mechanism is significantly different from the conventional BEC as no energy minimum exists in this region. To reveal characteristic features of the observed accumulation effect, time-resolved BLS measurements for both the narrow (50 μm) and the broad (500 μm) pumping areas were performed (see Fig. 12d-f). A pronounced population peak is formed at the magneto-elastic crossover at $q_\parallel = 0.8 \times 10^5 \, \text{rad} \, \text{cm}^{-1}$ in the case of the wide pumping area (see Fig. 12e). On the contrary, in compliance with our previously reported results [81], no accumulation is visible in this spectral region in the case of the narrow pumping area (see Fig. 12d). During further time evolution, the magnons occupy energy states around the global energy minimum and tend to form a Bose-Einstein condensate. The enhancement of the BEC formation in a freely evolving magnon gas leads to the appearance of the BEC's density peak just after the termination of the pumping pulse [15]. Remarkably, the magneto-elastic peak emerges at pumping powers at least ten times smaller than the threshold of the BEC formation (see Fig. 12f). Nevertheless, with the increase of the pumping power (and consequently of the BEC density), further growth in the population of the magneto-elastic peak is suppressed (see black and red lines in Fig. 12f).

Obviously, the mechanism, which is responsible for the observed accumulation, must significantly differ from that in the conventional BEC, as no energy minimum exists in this region (see the dispersion in Fig. 11b). To understand the observed phenomenon, we should consider the physics of the scattering processes leading to the thermalization of the parametrically driven magnon gas. The excitation of parametric magnons (see Fig. 11b) populates the isofrequency surface at a half of the pumping frequency. The essential part of these magnons are in the exchange region of the spectrum, where the four-magnon scattering processes dominate over other types of interactions. This distributes parametric magnons over the entire wavevector space in such a way that the magnon number flux is mostly directed toward lower frequencies. The fastest track for the exchange magnons toward the bottom of their spectrum is through the lowest magnon mode with wavevectors parallel to the external bias field [83]. During the thermalization process, magnons reach the area of hybridization between the lowest magnon mode and the transverse phonon branch, gradually converting into hybrid magnon-phonon quasiparticles. As phonons are very linear particles and does not support a nonlinear scattering, these quasiparticles cannot scatter further down by frequency along the phonon branch. In the same time, their scattering ability to the other (higher by frequency) magnon and magnon-phonon branches is also limited, as the scattering rate is proportional to the populations of the interacting quasiparticle's spectral groups. This bottleneck in the scattering efficiency leads to building up a high population of the hybridization region of the spectrum. It is necessary to note that the bottleneck accumulation of hybrid magnetoelastic bosons is exclusively related to the variable scattering interaction of quasiparticle hybrids and, thus, should not be mixed with a "bottleneck" in a relaxation chain from an externally excited quasiparticle state to the phonon bath. In our case, the latter effect is a prerequisite for an increase in the density of externally pumped gaseous magnons above the thermal level. The magnon-phonon bottleneck accumulation phenomenon is not unique for this particular system and can occur in any multicomponent gas mixture of interacting quasiparticles with significantly different scattering amplitudes. Furthermore, unlike BEC, the accumulated magneto-elastic bosons possess a nonzero group velocity, making them promising data carriers in prospective magnon spintronics [24] circuits.

### 4.3. Heat gradient-induced magnon supercurrent

By analogy to superconductivity and ultracold atomic systems, quasiparticle BECs reveal the macroscopic quantum phenomena like supercurrents, Josephson effects, second sound, and Bogoliubov waves. In order to be observed, these effects require a coherent interaction between two spatially separated BECs or a presence of a phase gradient in the condensate's wavefunction. Starting from the discovery of a magnon BEC, the intense work is being performed to control the magnon BEC localization and phase. One of the first demonstrations of two separated magnon BECs was done in Ref. [82]. By making use of different excitation geometries – namely, a thin wire and a microstrip, the different spatial localization of the condensate was achieved. Two spatially separated condensate clouds were simultaneously created in a magnetic film in the case of the 500 $\mu$m-wide microstrip resonator. However, no interaction between condensates was observed that time.

In the geometry of in-plane magnetization conventional for the magnon BEC observation (see e.g. Fig. 11a), the strong magnetic dipole-dipole interaction results in two symmetric minima in the frequency spectrum. Consequently, a system of two BECs with opposite wavevectors is created. These condensates can form a specific interference pattern, which can be revealed by means of the microfocused BLS technique [83]. The mutual coherence of two condensates leads to the establishment of a standing wave with distinct signatures of phase slips in the form of quantized vortices.

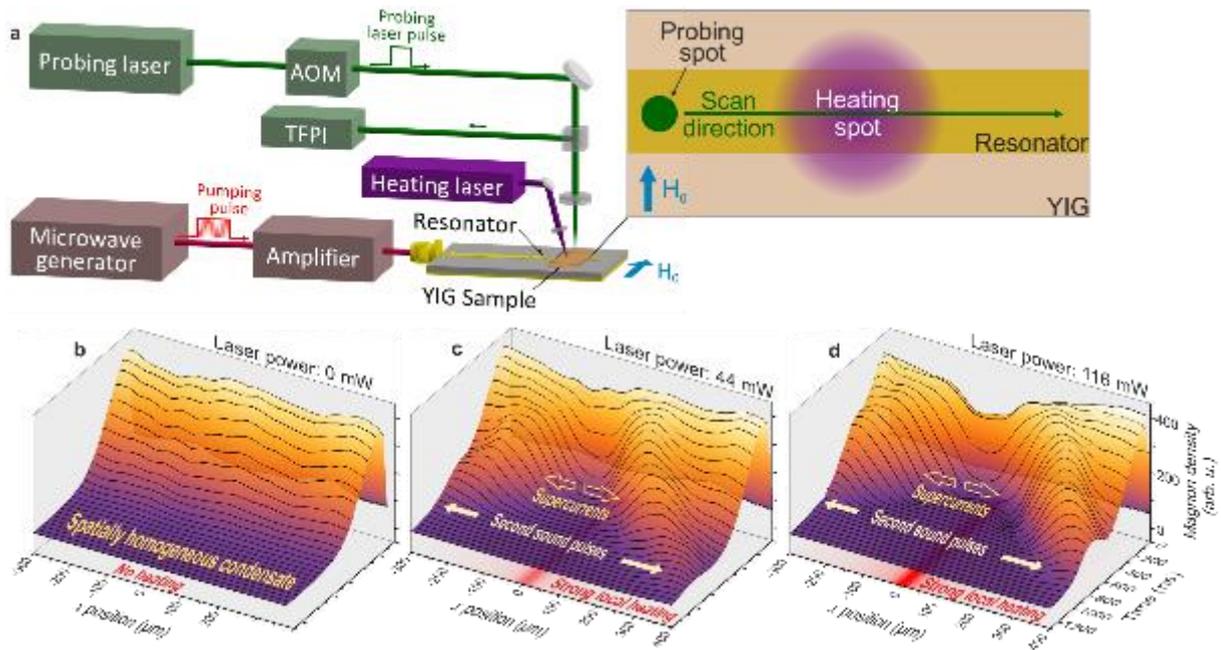

**Fig. 13. Observation of heat gradient induced magnon supercurrent and Bogoliubov waves.** (a) Schematic illustration of the experimental setup. A probing green laser beam of 532 nm wavelength is focused onto a YIG-film sample, which is fixed on top of a pumping microstrip resonator driven by powerful microwave pulses. The probing beam is chopped by an acousto-optic modulator (AOM) to reduce parasitic heating of the sample. The light and microwave pulses are synchronized allowing for optical observation of the after-pumping evolution of a magnon BEC. The light scattered by magnons is directed to a tandem Fabry-Pérot interferometer for intensity-, frequency-, and time-domain analysis. A blue laser of 405 nm wavelength is used for local heating of the sample and is focused into a 80 μm spot in the middle point of the resonator. The blue laser source and the YIG sample are mounted on a single movable stage to hold unchanged a predefined position of the heated area on YIG-film surface in the process of sample motion. The motion of the stage relative to the focal green spot allows for the probing of the magnon gas density in different points of the sample. The inset schematically shows a magnified view of the scan area. In order to ensure constant pumping conditions and, thus, spatially uniform density distribution of the magnon BEC in the probing direction, the scan is performed in x-direction along the microstrip resonator (across the bias magnetic field H). Spatio-temporal diagrams of the BEC propagation measured at different heating laser powers – (b) 0 mW, (c) 44 mW, and (d) 116 mW [84].

In order to explore further the coherence of a magnon BEC and make use of it to develop novel wave-based computing schemes, one needs a mechanism of local control over condensate's phase. One of such methods is to create a localized frequency shift of the magnon dispersion by the local heating [16]. The experiment was done in a classical geometry suitable for the observation of a magnon BEC magnetization in a 6.7 $\mu$m-thick YIG film (see Fig. 11a). The temporal evolution of the magnon BEC formed in a parametrically populated magnon gas was studied by means of time- and wavevector-resolved BLS spectroscopy. The thermal gradient in the condensate was created by a focused probing laser beam. To achieve the control over the heating, the laser beam was chopped by an acousto-optical modulator. It has been found that the local heating at the focal point of a probing laser beam leads to the excessive decay of a magnon BEC, which is associated with the outflow of condensed magnons driven by a thermal gradient. The magnon BEC flow occurs in the opposite to a seemingly obvious direction toward a potential well created in the hot region, which confirms the phase-dependent nature of the observed phenomenon – magnon supercurrent [16].

The first observation of a magnon supercurrent [16] was rather indirect. In the following experiment (see Fig. 13a), the heating and probing beams were separated in order to directly investigate the propagation of magnon supercurrent pulses [84]. In this experiment, the heating was provided by a continuous 405 nm laser, which is well separated from the 532-nm wavelength of a probing laser and also well absorbed by an YIG film. By shifting the detection point across and outside the heated area, the spatio-temporal diagrams of the BEC propagation were measured (see Fig. 13b-d). In all the cases, the formation of a BEC occurs at the end of the pumping pulse [15]. In the case of no additional heating, the BEC's decay is uniform around the scanned area (see Fig. 13b). When the heating is on, a clearly visible well in the magnon population is formed (see Fig. 13c) and becomes more pronounced with increase of the heating power (see Fig. 13d). In the same time, we observe magnons, which are being pushed out by magnon supercurrents and form distinct peaks at the sides of the heated area after ~200 ns. It is remarkable that these peaks do not stay at the edge of the heating area (as it is expected from the supercurrent-driven transport), but start to propagate with an initial group velocity of 410 m/s. As the condensate propagates, its population decreases, since magnons naturally decay. The group velocity of the BEC peaks also decreases to 361 m/s after ~1200 ns of the propagation time. This phenomenon can be understood as the excitation of a new type of second sound: Bogoliubov waves in the magnon BEC condensate. As the population of the BEC decreases, the group velocity $c_S$ decreases as well, as a result of the dependence of the dispersion of Bogoliubov waves on the condensate's population $N_C : c_S \propto \sqrt{N_C(t)}$ (which follows from the characteristics of magnon-magnon interactions [85]). Despite the decay and a decrease of the group velocity, the propagation was detected over the macroscopic distances up to more than 600 $\mu$s. The discovered magnon second sound differs from the second sound in dielectrics, in which the phonons can be described in terms of their occupation numbers only, not taking their phases into account. It also differs from the second sound in superfluid $^4$He and in the BEC of diluted atomic systems, where the wave function describes the distribution of real atoms and not of quasiparticles, as in the case of magnons. From a practical point of view, the three observed phenomena: (i) the transition from the supercurrent-type to the second-sound-type propagation regime, (ii) the excitation of second-sound-pulses, and (iii) the possibility of a long-distance spin-transport in the magnon BEC, has already paved a way for the application of magnon macroscopic quantum states for low-loss data transfer and information processing in perspective magnon spintronic devices. There are no doubts that the magnon second sound requires further detailed experimental and theoretical investigations. For example, the excitation of a second sound waves in a dense magnon gas was recently realized by using local magnetic fields generated by current pulses in microstructured wires, as has been reported in Ref. [86].

**Acknowledgements**

We acknowledge gratefully the financial support from the European Research Council (ERC Starting Grant 678309 MagnonCircuits and ERC Advanced Grant 694709 SuperMagnonics) and the Deutsche Forschungsgemeinschaft (DFG project B04 of the Research Unit TRR 173 "Spin+X").